\documentclass[a4paper,11pt]{article}

\usepackage{jcappub} % JCAP manuscript style
\usepackage{lineno}

\usepackage{graphicx}
\usepackage{dcolumn}
\usepackage{bm}
\usepackage{amssymb}
\usepackage{latexsym}
\usepackage{booktabs}
\usepackage{amsmath}
\usepackage{mathrsfs}
\usepackage{multirow}
\usepackage{url}
\usepackage{hyperref}
\usepackage{footnote}
\usepackage{float}
\usepackage{placeins}
\usepackage[font=small,labelfont=bf,labelsep=period]{caption}
\usepackage{threeparttable}
\usepackage{color}
\usepackage{array}
\usepackage{enumerate}
\usepackage{adjustbox}

\usepackage{makecell}
\usepackage{diagbox}
\usepackage{epstopdf}
\usepackage{epsfig}
\usepackage{longtable}
\usepackage{supertabular}
\usepackage{algorithm}
\usepackage{pifont}
\usepackage{algorithmic}
\usepackage{changepage}
\usepackage{setspace}

\begin{document}

%\title{An attempt to parametrize early- and late-time solutions: cosmic age}
\title{GPAge: a Global Parameterization based on cosmic Age for some unified early-time and late-time solutions to the Hubble tension}

\author[a,b]{Jia-Le Ling,}
\emailAdd{lingjiale@itp.ac.cn}

\author[a,c\ast]{Shao-Jiang Wang}
\emailAdd{schwang@itp.ac.cn (corresponding author)}

\affiliation[a]{Institute of Theoretical Physics, Chinese Academy of Sciences (CAS), Beijing 100190, China}
\affiliation[b]{School of Physical Sciences, University of Chinese Academy of Sciences (UCAS), Beijing 100049, China}
\affiliation[c]{Asia Pacific Center for Theoretical Physics (APCTP), Pohang 37673, Korea}

\abstract{The over $5\sigma$ Hubble tension continues to motivate modifications of the standard $\Lambda $-cold-dark-matter ($\Lambda \rm CDM$) model. The large number of proposed resolutions for the Hubble tension, broadly divided into early- and late-time solutions, motivates phenomenological parametrizations to constrain the viable solution space. Recently, there is a growing interest in unifying early- and late-time solutions. In this work, we extend the original Parametrization based on cosmic Age (PAge) to cover both the early and late Universe. Specifically, we incorporate the radiation-dominated epoch at the background level and complete the perturbation sector within a unified-dark-fluid framework. The resulting parametrization can accurately reproduce target cosmological models at both the background and perturbation levels. Confronting this global PAge with cosmological observations, we find that it improves the best fit relative to $\Lambda \rm CDM$, but does not raise $H_0$ significantly enough to alleviate the Hubble tension. Our parametrizations indicate that varying the early-time matter-like density, even together with a flexible late-time expansion history, is insufficient to resolve the Hubble tension. These results suggest that resolving the Hubble tension requires pre-recombination physics beyond a simple retuning of the matter-like density.
}
\maketitle

\section{Introduction}\label{sec:introduction}

The over $5\sigma $ discrepancy in the Hubble constant $H_0$ between the local distance-ladder measurement by the SH0ES (Supernovae and $H_0$ for the Equation of State of dark energy) collaboration~\cite{Riess:2021jrx} and the value inferred from Planck 2018 cosmic microwave background (CMB) observations assuming Lambda-cold-dark-matter ($\Lambda \rm CDM$)~\cite{Planck:2018vyg} has become the prominent challenge in modern cosmology~\cite{DiValentino:2016hlg,Karwal:2016vyq,Bernal:2016gxb,Verde:2019ivm,Vagnozzi:2019ezj,DiValentino:2020zio,Abdalla:2022yfr,Verde:2023lmm,CosmoVerseNetwork:2025alb,Cai:2026swf}. This discrepancy, referred to as the Hubble tension, has not yet been conclusively explained by known systematic effects~\cite{Zhang:2017aqn,Freedman:2021ahq,Kenworthy:2022jdh,Wojtak:2022bct,Riess:2025chq}, and persists across a variety of independent late- and early-Universe probes~\cite{Cuceu:2019for,ACT:2023kun,SPT-3G:2025bzu,DESI:2025zgx,Jensen:2025aai,Bhardwaj:2025kbw,Kourkchi:2020iyz,Pesce:2020xfe,H0LiCOW:2019pvv,Hotokezaka:2018dfi,Schoneberg:2019wmt}. The Hubble tension, together with other cosmological tensions~\cite{Bull:2015stt,Perivolaropoulos:2021jda,Abdalla:2022yfr,Arendse:2019hev,Aylor:2018drw,Wang:2026kor,Huang:2024gfw,Huang:2024erq,DiValentino:2020vvd,DiValentino:2022fjm,Poulin:2022sgp,Poulin:2024ken,Nguyen:2023fip,Pantos:2026koc}, may point toward new physics beyond the standard cosmological model, i.e., $\Lambda \rm CDM$~\cite{DiValentino:2020zio,DiValentino:2021izs,Hu:2023jqc, Perivolaropoulos:2021jda, Knox:2019rjx, Schoneberg:2021qvd, Bull:2015stt, Vagnozzi:2019ezj, Kamionkowski:2022pkx, Shah:2021onj,Schoneberg:2026eys, Schoneberg:2026vaf,Abdalla:2022yfr,Poulin:2023lkg,Vagnozzi:2021gjh,Vagnozzi:2023nrq,CosmoVerseNetwork:2025alb}.
However, there is still no consensus on the solution to the Hubble tension. The Hubble tension can be recast as an early-Universe sound-horizon $r_{\mathrm{s}}$ tension, where the CMB acoustic scale is related to the sound horizon through $r_{\mathrm{s}}=\theta_* d_{\mathrm{M}}/H_0$ and baryon acoustic oscillations (BAO) exhibit the degeneracy $H_0 r_{\mathrm{s}}\simeq{\rm const}$~\cite{Jiang:2024xnu,Bernal:2016gxb,Aylor:2018drw,Knox:2019rjx,Schoneberg:2019wmt}. Alternatively, it can be understood as a late-Universe tension in the absolute magnitude $M_{\rm B}$ of Type Ia supernovae (SNe) owing to the $M_{\rm B}$--$H_0$ degeneracy in the SNe distance ladders~\cite{Efstathiou:2021ocp,Camarena:2023rsd,Poulin:2024ken,Perivolaropoulos:2024yxv}.

Under this sound-horizon interpretation, the corresponding early-time solutions aim to modify pre-recombination physics and reduce the acoustic ruler while preserving the observed CMB structure~\cite{Aylor:2018drw,Knox:2019rjx}. Proposed mechanisms include modified recombination~\cite{Chluba:2023xqj,Chiang:2018xpn,Liu:2019awo,Sekiguchi:2020teg}, primordial magnetic fields~\cite{Jedamzik:2020krr,Vachaspati:2020blt,Jedamzik:2025cax}, additional relativistic species~\cite{Aloni:2021eaq,Blinov:2020hmc,Schoneberg:2022grr}, and nonstandard neutrino physics~\cite{Blinov:2019gcj,RoyChoudhury:2020dmd,Cyr-Racine:2013jua,Sakstein:2019fmf}. Among these possibilities, early dark energy (EDE) has received particular attention. EDE introduces a cosmological-constant-like component at early times ($z \gtrsim 3000$) that increases the pre-recombination expansion rate and subsequently dilutes rapidly~\cite{Poulin:2018cxd,Poulin:2018dzj,Smith:2019ihp,Niedermann:2020dwg,Du:2026qtq,Karwal:2016vyq,Agrawal:2019lmo,Lin:2019qug,Lin:2020jcb,Jiang:2021bab,Lin:2022phm,Cruz:2023lmn,Sakstein:2019fmf,Braglia:2020auw,Kamionkowski:2022pkx,Poulin:2023lkg,Niedermann:2019olb,Ye:2020btb,Karwal:2021vpk,Gonzalez:2020fdy,Freese:2021rjq,Jing:2024pvh,Bella:2026zuk,Niedermann:2021ijp,Niedermann:2021vgd,Eskilt:2023nxm}.
However, raising $H_0$ in EDE models tends to exacerbate the matter-perturbation linear-amplitude ($S_8$) tension. EDE models also face the coincidence problem, fine-tuning problem, and additional no-go arguments~\cite{Vagnozzi:2023nrq,Jedamzik:2020zmd,Krishnan:2020obg,Hill:2020osr}. 
Under the $M_\mathrm{B}$-tension interpretation, the corresponding late-time solutions aim to modify the local distance calibration or the late-time expansion history. Proposed mechanisms include variations in the effective Newtonian constant $G_{\rm eff}$~\cite{Marra:2021fvf,Ruchika:2023ugh} and an effective shift in the SN Ia absolute magnitude induced by late-time dark-energy dynamics~\cite{Yang:2021eud,Akarsu:2019hmw,Keeley:2019esp,DiValentino:2020naf,Alestas:2020zol,Li:2019yem,Li:2025eqh}. However, such late-time solutions are strongly constrained by the inverse distance ladder and various late-time no-go theorems~\cite{Huang:2024erq,Ling:2025lmw,Cai:2022dkh,Cai:2021weh,Heavens:2014rja,Guo:2018ans,Knox:2019rjx,Efstathiou:2021ocp,Tiwari:2026pzk,Addison:2017fdm,Alestas:2021xes}.

In light of the large variety of proposed cosmological models and the corresponding no-go arguments, phenomenological parametrizations provide a useful framework for narrowing the viable solution space~\cite{Ling:2025lmw,Du:2025csv,Sahni:2002fz, Visser:2003vq, Krishnan:2021dyb, Lemos:2018smw, Capozziello:2011tj, Escamilla:2023shf,Luongo:2024fww,Aviles:2012ay}. The $w_0w_a\rm CDM$ model is a widely used phenomenological parametrization of dynamical dark energy (DDE)~\cite{Chevallier:2000qy, Linder:2002et}. Notably, Dark Energy Spectroscopic Instrument (DESI) analyses have reported a $4.2\sigma$ preference for DDE~\cite{DESI:2025zgx}, motivating renewed discussion of the nature of DE~\cite{Yang:2024kdo,Cai:2025mas,Qiu:2025oop,Zhou:2026dti,Li:2026xaz,Zhang:2025dwu,Du:2025xes,Li:2025owk,Li:2024qso,Wang:2025znm,Ye:2024ywg,Tada:2024znt,Khoury:2025txd,Giare:2024smz,Lyu:2025nsd,Chen:2025wwn,Petri:2025swg,Guedezounme:2025wav,Jhaveri:2026bla,Yashiki:2025loj,Wolf:2025jed,Ferrari:2025egk,Ye:2024zpk,Dai:2026xau,Ren:2026jyw,Li:2026asg}.
Most existing phenomenological parametrizations have focused on the late-time Universe~\cite{Alam:2025epg,Wang:2025xvi,Ozulker:2025ehg,Lopez-Hernandez:2024osv,Dinda:2024ktd,Escamilla:2024fzq,Giare:2024gpk,Dinda:2024kjf,Krishnan:2021dyb,DiValentino:2020naf,Holsclaw:2010sk, Zhao:2017cud, Jiang:2024xnu,Zhang:2025bmk,Wang:2019vxv, Gomez-Vargas:2021zyl}. Extending them to the early Universe is more challenging, since CMB observations do not directly measure $H(z)$~\cite{Hojjati:2013oya,Moss:2021obd,Samsing:2012qx,Doran:2006kp}. This motivates a parametrization that can consistently describe both late- and early-time expansion. A natural starting point is the Parametrization based on cosmic Age (PAge) of matter-dominated era supplemented with a dark-energy era, which has been shown to accurately describe late-time expansion histories deep into matter era~\cite{Luo:2020ufj,Huang:2020mub,Cai:2021weh,Huang:2021tvo,Cai:2022dkh,Huang:2024erq,Ling:2025lmw,Jia:2026ktx}, making its extension toward the early Universe an attractive direction to explore.

In this work, we extend the original PAge framework to also include the radiation-dominated epoch to be a truely global parametrization at the background level. Since the background evolution alone does not uniquely determine cosmological perturbations~\cite{Gubitosi:2012hu,Hu:1998kj}, we complete the perturbation sector within a unified-dark-fluid (UDF) framework~\cite{Wang:2024rus,Su:2025ntt}, without imposing a model-dependent separation between DM and DE, as motivated by the dark-sector degeneracy and previous unified-dark-sector studies~\cite{Kunz:2007rk,vonMarttens:2019ixw,Chen:2025wwn,Kunz:2009yx,Petri:2025swg,Carneiro:2014uua,You:2026qzi,Kamenshchik:2001cp,Bento:2002ps,Bilic:2001cg,Zhang:2004gc,Scherrer:2004au,Bertacca:2008uf,Bertacca:2010ct,Arbey:2006it,Padmanabhan:2002sh,BeltranJimenez:2026tlr,Colistete:2007xi,Chavanis:2015eka,Wetterich:2001yw,Brilenkov:2013yva,Elkhateeb:2021atj,Kou:2025yfr,Arbey:2008gw,Elkhateeb:2018unf,SevillanoMunoz:2026jgk,Castelao:2026xif,Aviles:2011ak}.
We consider two data combinations built from CMB, DESI DR2 BAO, DES--Dovekie SNe, with and without SH0ES, to test whether this global PAge (GPAge) model can alleviate the Hubble tension. We also examine the resulting $S_8$ constraints and the late-time dynamics of the unified dark fluid, in light of recent DESI indications of DDE.

This work is organized as follows. In Sec.~\ref{sec:model}, we introduce our GPAge model and its unified-dark-fluid perturbative completion. In Sec.~\ref{sec:data-methodology}, we describe the cosmological data and inference methodology. In Sec.~\ref{sec:results-discussion}, we present the constraints and discuss their implications. Finally, we summarize our conclusions in Sec.~\ref{sec4}.

\section{Extending PAge to the early Universe}\label{sec:model}

\subsection{Original PAge parametrization }\label{sec:original PAge parametrization}

The original PAge model is based on the matter-dominated limit, $Ht=\frac{2}{3}$, which holds exactly in a pure matter-dominated Universe. Assuming that the short radiation-dominated epoch can be neglected, the original PAge parametrizes the late-time departure from matter domination induced by DE with a quadratic expansion in cosmic time~\cite{Huang:2020mub}. This gives

\begin{equation}
    Ht-\frac{2}{3}=\left[p_{\rm age}-\frac{2}{3}(1+\eta_1)\right]\left(\frac{t}{t_0}\right)+\frac{2}{3}\eta_1\left(\frac{t}{t_0}\right)^2,
    \label{eq:page-original}
\end{equation}
where $t$ is the cosmic age, $t_0$ is the current cosmic age, $p_{\rm age}\equiv H_0t_0$, and $\eta_1$ characterizes the late-time deviation from matter domination. It is related to the present-day deceleration parameter $q_0$ through
\begin{equation}
    \eta_1=1-\frac{3}{2}p_{\rm age}^2(1+q_0).
    \label{eq:2.2}
\end{equation}
The expansion history can then be obtained by solving
\begin{equation}
    H=-\frac{1}{1+z}\frac{{\rm d}z}{{\rm d}t},
\end{equation}
together with Eq.~\eqref{eq:page-original} for the combination $H_0t$ as a function of $z$, $p_{\rm age}$, and $\eta_1$~\cite{Cai:2021weh,Cai:2022dkh,Huang:2024erq}.

\subsection{PAge across the radiation--matter transition}
\label{sec:page-radiation-extension}

Neglecting the short radiation-dominated epoch is adequate for the original late-time PAge parametrization. However, extending the parametrization to pre-recombination physics requires radiation to be incorporated explicitly. Previous PAge-like unified-dark-fluid constructions treated radiation separately and applied the PAge parametrization only to the remaining components~\cite{Wang:2024rus,Su:2025ntt}. Here, instead, we incorporate the radiation--matter transition directly into the PAge background. We replace the matter-dominated baseline $Ht=2/3$ of the original PAge model by the two-component co-evolution relation
\begin{equation}
Ht\equiv\frac23 f, \qquad 
f=\frac{(\tau/\tau_*+1)(\tau/\tau_*+3)}{(\tau/\tau_*+2)^2},
\label{eq:2.3}
\end{equation}
which follows from the exact radiation--matter background solution
\begin{equation}
a(\tau)=a_{\rm eq}\left[\left(\frac{\tau}{\tau_*}\right)^2+2\left(\frac{\tau}{\tau_*}\right)\right],
\qquad
\tau_*=\frac{\tau_{\rm eq}}{\sqrt{2}-1}.
\label{eq:2.4}
\end{equation}
Here, $\tau_{\rm eq}$ and $a_{\rm eq}$ denote the conformal time and scale factor at matter--radiation equality, respectively. 

Following the original PAge construction and its higher-order extension~\cite{Huang:2021aku}, we introduce the additional coefficients $\eta_2$ and $\eta_3$ to further capture the late-time deviation, and define
\begin{equation}
x\equiv H_0t,
\qquad
u\equiv\frac{x}{p_{\rm age}},
\qquad
f_0\equiv f(\tau_0),
\qquad
y\equiv H_0\tau,
\qquad
y_*\equiv H_0\tau_*,
\qquad
y_0\equiv H_0\tau_0.
\end{equation}
The extended PAge relation of our GPAge model is then written as
\begin{align}\label{eq:strict-page-E}
Ht-\frac{2}{3}f=&\left[p_{\rm age}-\frac{2}{3}f_0(1+\eta_1+\eta_2+\eta_3)\right]u\nonumber\\
&+\frac{2}{3}f_0(\eta_1+2\eta_2+3\eta_3)u^2-\frac{2}{3}f_0(\eta_2+3\eta_3)u^3+\frac{2}{3}f_0\eta_3u^4 .
\end{align}
The full expansion history is obtained by solving a system of first-order differential equations
\begin{equation}
x\frac{H}{H_0}=\frac{{\rm d}\ln a}{{\rm d}\ln x},
\qquad
\frac{{\rm d}\tau}{{\rm d}t}=\frac{1}{a},
\end{equation}
subject to
\begin{align}
a(t_0)=1, \qquad \tau(t=0)=0,
\end{align}
where the contribution to the conformal time from the inflationary epoch is neglected. Analogously to Eq.~\eqref{eq:2.2}, the present-day deceleration parameter
is
\begin{equation}
q_0=-1+\frac{2f_0}{3p_{\rm age}^2}(1-\eta_1)-\frac{2f_{x,0}}{3p_{\rm age}},
\qquad
f_{x,0}\equiv\left.\frac{{\rm d}f}{{\rm d}x}\right|_{x_0}=\frac{2}{H_0\tau_*(\tau_0/\tau_*+2)^3}.
\label{eq:page-q0}
\end{equation}

At this stage, the background construction of our GPAge model involves the quantities
\begin{equation}
\boldsymbol{\theta}_{\rm PAge}=\left\{p_{\rm age},\eta_1,\eta_2,\eta_3,y_*,y_0\right\}.
\label{eq:2.8}
\end{equation}
The parameters $p_{\rm age}$ and $y_0$ characterize global, integrated properties of the expansion history, while $y_*$ controls the radiation--matter transition. The coefficients $\eta_1,\eta_2,\eta_3$ mainly control the late-time deformation, since their contributions are suppressed in the early-time limit $u\to0$. In what follows, we organize the GPAge family according to the number of
independent GPAge coefficients,
\begin{equation}
\begin{aligned}
{\rm PAge_2}:&\quad \{p_{\rm age},\eta_1\}, &&\eta_2=\eta_3=0,\\
{\rm PAge_3}:&\quad \{p_{\rm age},\eta_1,\eta_2\}, &&\eta_3=0,\\
{\rm PAge_4}:&\quad \{p_{\rm age},\eta_1,\eta_2,\eta_3\}.
\end{aligned}
\label{eq:page-hierarchy}
\end{equation}
Although $y_*$ and $y_0$ enter the background construction, they are not independent model parameters. For each set of GPAge coefficients, they are determined by the two self-consistency conditions discussed in Sec.~\ref{sec:shooting}.

\begin{figure}[!htbp]
\centering
\includegraphics[width=\textwidth]{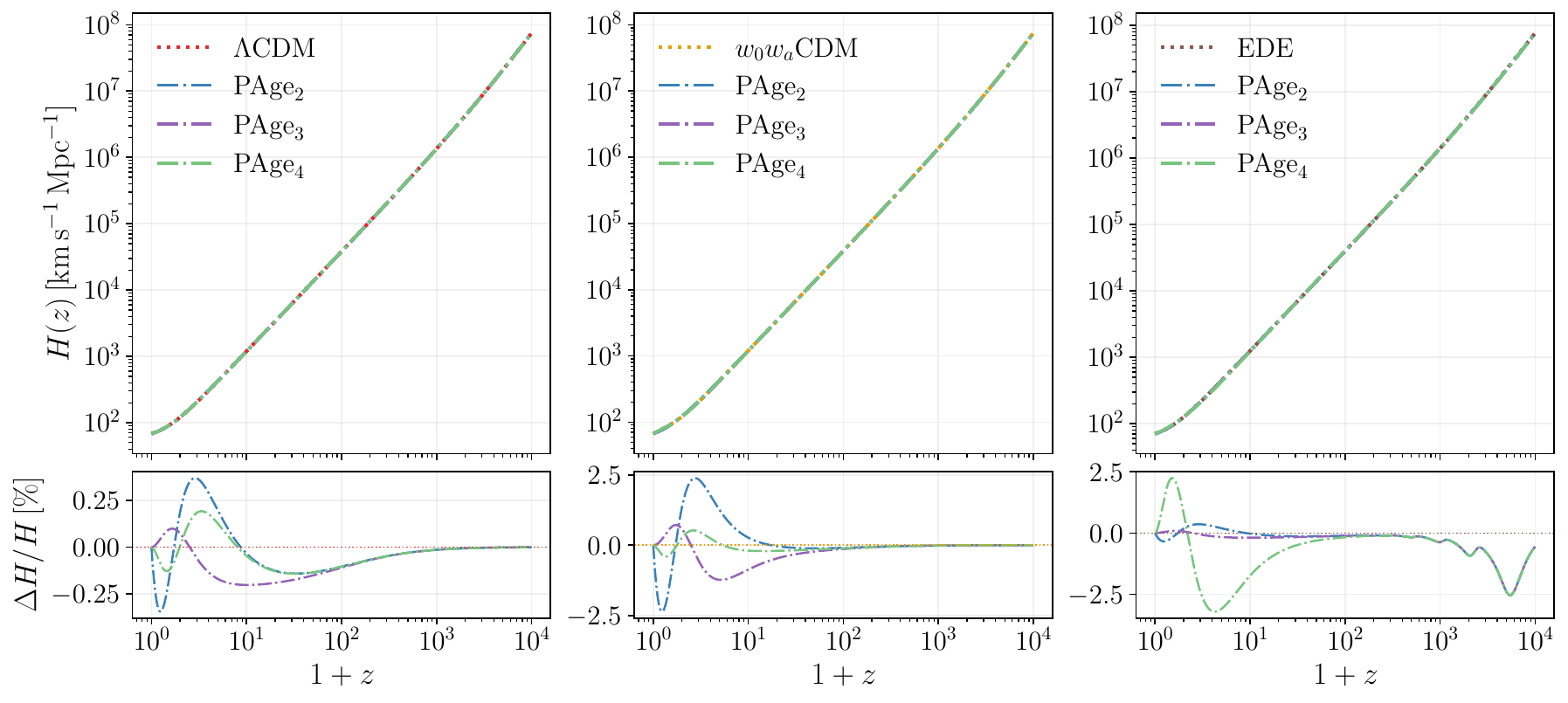}
\caption{Background mappings of representative $\Lambda \rm CDM$, $w_0w_a \rm CDM$, and EDE expansion histories using $\mathrm{PAge}_2$, $\mathrm{PAge}_3$, and $\mathrm{PAge}_4$. The upper panels show $H(z)$, and the lower panels show the corresponding relative residuals, normalized to the corresponding target models.}
\label{fig:background-mapping}
\end{figure}

Before confronting the models with cosmological data, we examine their ability to reproduce target expansion histories. Fig.~\ref{fig:background-mapping} shows the GPAge mappings of $\Lambda \rm CDM$, $w_0w_a \rm CDM$, and EDE backgrounds. For $\Lambda$CDM, we adopt the posterior mean parameters of CMB+DESI+DES-Dovekie reported in Table~\ref{table:param_values}. For the $w_0w_a \rm CDM$, we adopt the posterior-mean parameters from the DESI+CMB+DESY5 analysis reported in Table V of Ref.~\cite{DESI:2025zgx},
\begin{align}
H_0=66.74~{\rm km\,s^{-1}\,Mpc^{-1}}, \qquad w_0=-0.752, \qquad w_a=-0.86 .
\end{align}
For EDE, we adopt the posterior mean parameters of the $n=3$
axion-EDE model reported in Table~II of
Ref.~\cite{Poulin:2018cxd},
\begin{align}
H_0=70.6~{\rm km\,s^{-1}\,Mpc^{-1}}, \qquad f_{\rm EDE}=0.05, \qquad \log_{10}a_c=-3.737 .
\end{align}
For each target model, the GPAge mapping is defined by matching a selected set of physical quantities. Denoting quantities evaluated in the target cosmology by the superscript ``tar'', we impose
\begin{equation}
\setlength{\jot}{2pt}
\begin{aligned}
&\mathrm{PAge}_2:\quad
p_{\rm age}^{\rm PAge}=p_{\rm age}^{\rm tar},\;
\lim_{a\to0}a^3\rho_{\rm m}^{\rm early,PAge}
=\lim_{a\to0} a^3\rho_{\rm m}^{\rm early,tar},\\
&\mathrm{PAge}_3:\quad
p_{\rm age}^{\rm PAge}=p_{\rm age}^{\rm tar},\;
q_0^{\rm PAge}=q_0^{\rm tar},\;
\lim_{a\to0}a^3\rho_{\rm m}^{\rm early,PAge}
=\lim_{a\to0}a^3\rho_{\rm m}^{\rm early,tar},\\
&\mathrm{PAge}_4:\quad
p_{\rm age}^{\rm PAge}=p_{\rm age}^{\rm tar},\;
q_0^{\rm PAge}=q_0^{\rm tar},\;
100\theta_s^{\rm PAge}=100\theta_s^{\rm tar},\;
\lim_{a\to0}a^3\rho_{\rm m}^{\rm early,PAge}
=\lim_{a\to0}a^3\rho_{\rm m}^{\rm early,tar}.
\end{aligned}
\label{eq:page-mapping}
\end{equation}
Here, $\rho_{\rm m}$ is the matter density and $\theta_s$ is the angular size of the sound horizon at last scattering. The early-time matter density $\rho_{\rm m}^{\rm early,PAge}$ is defined below in Eq.~\eqref{eq:2.28}. We first calculate $p_{\rm age}^{\rm tar},~q_0^{\rm tar},~\lim_{a\to0} a^3\rho_{\rm m}^{\rm early,tar},~100\theta_s^{\rm tar}$ in the target model and use Eq.~\eqref{eq:page-mapping} to determine the PAge parameters.

\begin{figure}[!htbp]
\centering
\includegraphics[width=0.8\textwidth]{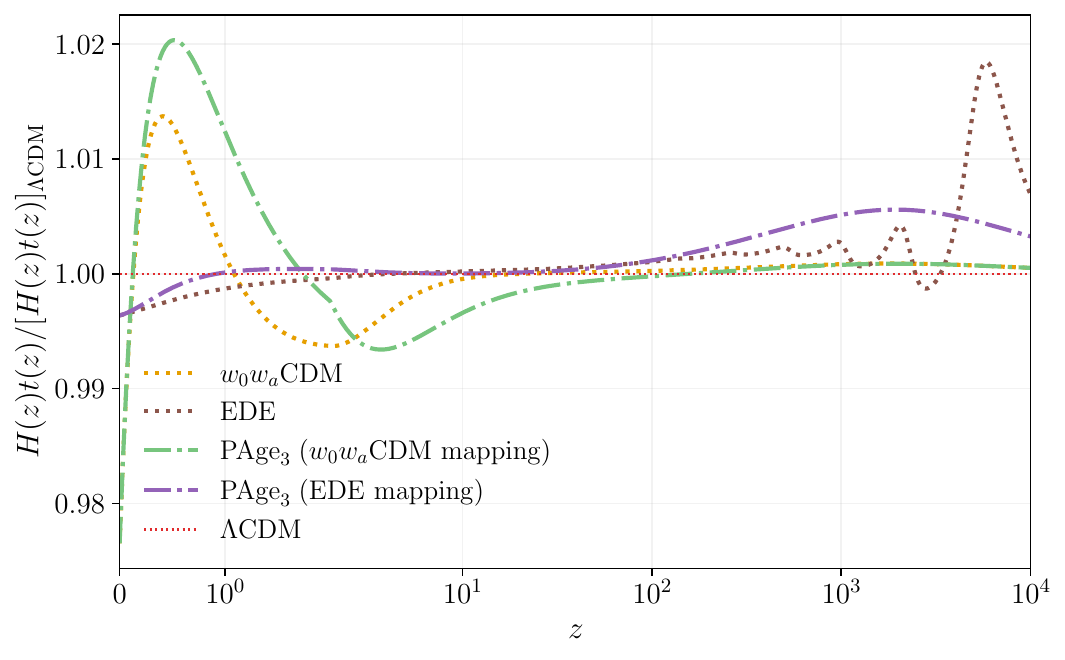}
\caption{$H(z)t(z)$ for the representative $w_0w_a \rm CDM$ and EDE backgrounds and their $\mathrm{PAge}_3$ mappings, normalized to the \(\mathcal D_0\) posterior-mean $\Lambda \rm CDM$.}
\label{fig:clock-mapping}
\end{figure}

As shown in Fig.~\ref{fig:background-mapping}, $\rm PAge_3$ already captures the representative $\Lambda \rm CDM$, $w_0w_a \rm CDM$, and EDE backgrounds very well with deviation less than $\sim 2\%$. The main residual feature appears at high redshift, where the EDE background develops a localized peak. Adding the higher-order coefficient $\eta_3$ in $\rm PAge_4$ yields little further improvement in reproducing these expansion histories. To understand the origin of the high-redshift mismatch in EDE, we turn to $H(z)t(z)$. As shown in Fig.~\ref{fig:clock-mapping}, $\rm PAge_3$ reproduces the EDE background well at low redshift. At high redshift, however, the localized energy injection characteristic of EDE is not fully captured. Although the GPAge mapping also exhibits an early-time enhancement, its amplitude remains significantly smaller than that of the axion-EDE.

This mismatch originates from Eq.~\eqref{eq:2.3}, where the early-time background is restricted to the analytic radiation--matter solution controlled by $y_*$. Since EDE contains an additional early-time component, a closer mapping to the EDE background could in principle be achieved by introducing more freedom into Eq.~\eqref{eq:2.3}. However, reproducing the rapid appearance and subsequent disappearance of EDE would require an additional localized feature, such as a Gaussian-like bump. Such a modification would be rather ad hoc and lack a clear physical motivation, and is therefore not included in the GPAge parametrization considered here.

\FloatBarrier

\subsection{Unified-dark-fluid perturbations}\label{sec:udf-completion}

Specifying the background expansion history does not uniquely determine the evolution of perturbations~\cite{Gubitosi:2012hu,Hu:1998kj}. To confront the PAge background with CMB observations, we therefore complete its perturbation sector within a unified-dark-fluid (UDF) framework~\cite{Hu:1998kj,Kodama:1984ziu}.

We begin by decomposing the Friedmann equation into the density of the unified dark fluid (UDF) $\rho_{\rm df}$ and the visible sector $\rho_{\rm visible}$,
\begin{equation}
3M_{\rm Pl}^2 H_{\rm PAge}^2=\rho_{\rm df}+\rho_{\rm visible},
\label{eq:udf-density}
\end{equation}
where $M^2_{\rm Pl}\equiv1/(8\pi G)$ and $H_{\rm PAge}$ is defined in Eq.~\eqref{eq:strict-page-E}. The visible sector consists of baryons $\rho_b$, photons $\rho_\gamma$, and neutrinos $\rho_\nu$,
\begin{equation}
\rho_{\rm visible}=\rho_b+\rho_\gamma+\rho_\nu,\qquad
p_{\rm visible}=p_\gamma+p_\nu,
\label{eq:known-sector}
\end{equation}
where the baryon pressure is neglected. The UDF background evolution obeys the continuity equation. We obtain
\begin{equation}
\begin{aligned}
p_{\rm df}&=-\rho_{\rm df}-\frac{\dot{\rho}_{\rm df}}{3(\dot a/a)},\\
w_{\rm df}&=\frac{p_{\rm df}}{\rho_{\rm df}},\\
c_{\rm s, ad}^2&=\frac{\dot{p}_{{\rm df}}}{\dot{\rho}_{{\rm df}}}=w_{\rm df}-\frac{\dot{w}_{{\rm df}}}{3\frac{\dot{a}}{a}(1+w_{\rm df})}.
\end{aligned}
\label{eq:udf-background-quantities}
\end{equation}
Here, a dot denotes differentiation with respect to conformal time $\tau$, $p_{\rm df}$ is the pressure of the unified dark fluid, $w_{\rm df}=p_{\rm df}/\rho_{\rm df}$ is its equation-of-state (EOS) parameter, and $c_{\rm s, ad}^2$ is UDF adiabatic sound speed.

We work in the synchronous gauge with metric of the form,
\begin{equation}
\mathrm{d}s^2=a^2(\tau)\left[-\mathrm{d}\tau^2+(\delta_{ij}+h_{ij})\mathrm{d}x^i \mathrm{d}x^j\right],
\end{equation}
where $\tau$ is the conformal time and $h_{ij}$ is the metric perturbation. To avoid the unphysical oscillations that can arise in some UDF models~\cite{Sandvik:2002jz}, we allow for entropy perturbations and introduce the UDF rest-frame effective sound speed $c_{\rm s, eff}^2$~\cite{Hu:1998kj}. The pressure perturbation is written as
\begin{equation}
\delta p_{\rm df}=c_{\rm s, eff}^2\,\delta\rho_{\rm df}+3\mathcal{H}(1+w_{\rm df})\rho_{\rm df}\left(c_{\rm s, eff}^2-c_{\rm s, ad}^2\right)\frac{\theta_{\rm df}}{k^2},
\label{eq:pressure-perturbation}
\end{equation}
where $\mathcal{H}\equiv\dot a/a$ is the conformal Hubble parameter, $\theta_{\rm df}$ is the velocity divergence of the unified dark fluid, $k$ is the comoving wavenumber, and $c_{\rm s,eff}^2$ is the effective sound speed in the UDF rest frame. At linear order, the density contrast $\delta_{\rm df}\equiv\delta\rho_{\rm df}/\rho_{\rm df}$ and the velocity divergence $\theta_{\rm df}$ obey
\begin{equation}
\dot{\delta}_{\rm df}=
-(1+w_{\rm df})\left(\theta_{\rm df}+\frac{\dot h^i_i}{2}\right)
-3\mathcal H\left(c_{\rm s, eff}^2-w_{\rm df}\right)\delta_{\rm df}
-9\mathcal H^2\left(c_{\rm s, eff}^2-c_{\rm s, ad}^2\right)(1+w_{\rm df})\frac{\theta_{\rm df}}{k^2},
\end{equation}
and
\begin{equation}
\dot{\theta}_{\rm df}=-\mathcal H\left(1-3c_{\rm s, eff}^2\right)\theta_{\rm df}
+\frac{c_{\rm s, eff}^2}{1+w_{\rm df}}k^2\delta_{\rm df}
-k^2\sigma_{\rm df}.
\end{equation}
Here, $h^i_i$ denotes the scalar trace metric perturbation in synchronous gauge, and $\sigma_{\rm df}$ is the UDF anisotropic stress. In this work, we adopt
\begin{equation}
c_{\rm s, eff}^2=0, \qquad \sigma_{\rm df}=0,
\label{eq:udf-closure-full}
\end{equation}
so that the unified dark fluid can cluster efficiently. We impose adiabatic initial conditions,
\begin{equation}
\frac{\delta_{\rm df}}{1+w_{\rm df}}=\delta_b=\frac{3}{4}\delta_\gamma=\frac{3}{4}\delta_\nu, \qquad (k\tau\ll1).
\label{eq:udf-adiabatic-initial}
\end{equation}
For the backgrounds considered here, $w_{\rm df}$ remains above $-1$, so the perturbation equations do not encounter a phantom crossing and no PPF prescription is required~\cite{Fang:2008sn,Hu:2007pj}.

To discuss the $S_8$ tension~\cite{Pantos:2026koc,DiValentino:2020vvd} or a phantom-crossing behavior~\cite{DESI:2025zgx}, we first need to define effective DM and effective DE components within the unified dark fluid. However, this decomposition is subject to the dark degeneracy. Since DM and DE are probed through gravity and enter the gravitational field equations through their energy-momentum tensors, gravitational observables constrain only the total dark-sector energy-momentum tensor. Without additional assumptions, the DM and DE components therefore cannot be uniquely separated at either the background or perturbation level~\cite{Kunz:2007rk}. This degeneracy can be expressed more explicitly as follows,
\begin{equation}
\rho_{\rm df}=\rho_{{\rm de},1}+\rho_{{\rm dm},1}=\rho_{{\rm de},2}+\rho_{{\rm dm},2},
\qquad
p_{\rm df}=p_{{\rm de},1}+p_{{\rm dm},1}=p_{{\rm de},2}+p_{{\rm dm},2},
\label{eq:220}
\end{equation}
where $1$ and $2$ represent different cosmological models. Assuming that the DM component is pressureless in both models, $p_{{\rm dm},1}=p_{{\rm dm},2}=0$, we have $p_{\rm df}=w_{{\rm de},1}\rho_{{\rm de},1}=w_{{\rm de},2}\rho_{{\rm de},2}$, and hence
\begin{equation}
w_{{\rm de},1}=\frac{w_{{\rm de},2}\rho_{{\rm de},2}}{\rho_{{\rm de},2}+\rho_{{\rm dm},2}-\rho_{{\rm dm},1}}=\frac{p_{\rm df}}{\rho_{\rm df}-\rho_{{\rm dm},1}}.
\label{eq:dark-degeneracy-eos}
\end{equation}
If model $1$ and model $2$ represent the same model viewed from two different perspectives, Eq.~\eqref{eq:220} holds exactly, and the difference between their inferred dark-energy equations of state can give rise to an apparent phantom crossing~\cite{Khoury:2025txd,Guedezounme:2025wav}. If, instead, models $1$ and $2$ are genuinely different cosmological models whose best-fit background evolutions do not coincide, Eq.~\eqref{eq:220} no longer holds exactly. In this case, applying Eq.~\eqref{eq:dark-degeneracy-eos} may lead to a spurious phantom-crossing with model mismatch, as discussed in Refs.~\cite{Kunz:2015oqa,Wang:2026wrk,Wang:2025znm}. 

In our UDF framework, only $\rho_{\rm df}$ and $p_{\rm df}$ are specified. First, to discuss the phantom-crossing behavior with DESI, we fix a reference CDM density $\rho_{\rm dm,1}=\rho_{\rm cdm}^{\Lambda \rm CDM}$ in Eq.~\eqref{eq:dark-degeneracy-eos} and define the effective EOS of DE,
\begin{equation}
w_{\rm de}^{\rm eff}=\frac{p_{\rm df}}{\rho_{\rm df}-\rho_{\rm cdm}^{\Lambda \rm CDM}}.
\label{eq:effective-de-eos}
\end{equation}
For the $S_8$ analysis, we fix $w_{{\rm de},1}=-1$ in Eq.~\eqref{eq:dark-degeneracy-eos} to get effective DM density,
\begin{equation}
\rho_{\rm dm}^{\rm eff}=\rho_{\rm df}-\frac{p_{\rm df}}{w_{\rm de,1}}=\rho_{\rm df}+p_{\rm df},
\label{eq:2.25}
\end{equation}

\subsection{Self-consistency conditions}
\label{sec:shooting}

In our GPAge, the parameters $y_*$ and $y_0$ are treated as auxiliary shooting variables rather than independent model parameters, and are fixed by imposing two self-consistency
conditions.

First, for generic $y_*$ and $y_0$, the residual UDF in Eq.~\eqref{eq:udf-density} may contain a spurious dark-radiation-like component at high redshift. While radiation-like at the background level, its perturbations still obey the UDF closure with $c_{\rm s,eff}^2=0$. We therefore shoot for an early-time PAge background satisfying
\begin{equation}
\frac{\rho_{\rm df}}{\rho_{\rm visible}}\to 0,\qquad
\frac{3M^2_{\rm Pl}H_{\rm PAge}^2}{\rho_{\rm visible}}\to 1\qquad
(u\to0),
\label{eq:rho}
\end{equation}
to eliminate the unphysical dark-radiation-like component. With Eq.~\eqref{eq:rho}, taking the early-time limit
$t\to0$ in Eq.~\eqref{eq:strict-page-E} gives
\begin{equation}
\rho_{\rm m}^{\rm early}(a)=\frac{2H_0}{a\,y_*}\sqrt{\frac{3\left[\rho_\gamma(a)+\rho_\nu(a)\right]}{8\pi G}},
\qquad a\to0 .
\label{eq:2.28}
\end{equation}
Thus, once the radiation abundance is fixed in Eq.~\eqref{eq:rho}, $y_*$ directly controls the effective early-time matter density. Although $y_*$ is determined by the shooting procedure rather than sampled independently, its self-consistent value varies with the PAge coefficients to adjust the early-matter density.

\begin{figure}[!htbp]
\centering
\includegraphics[width=\textwidth]{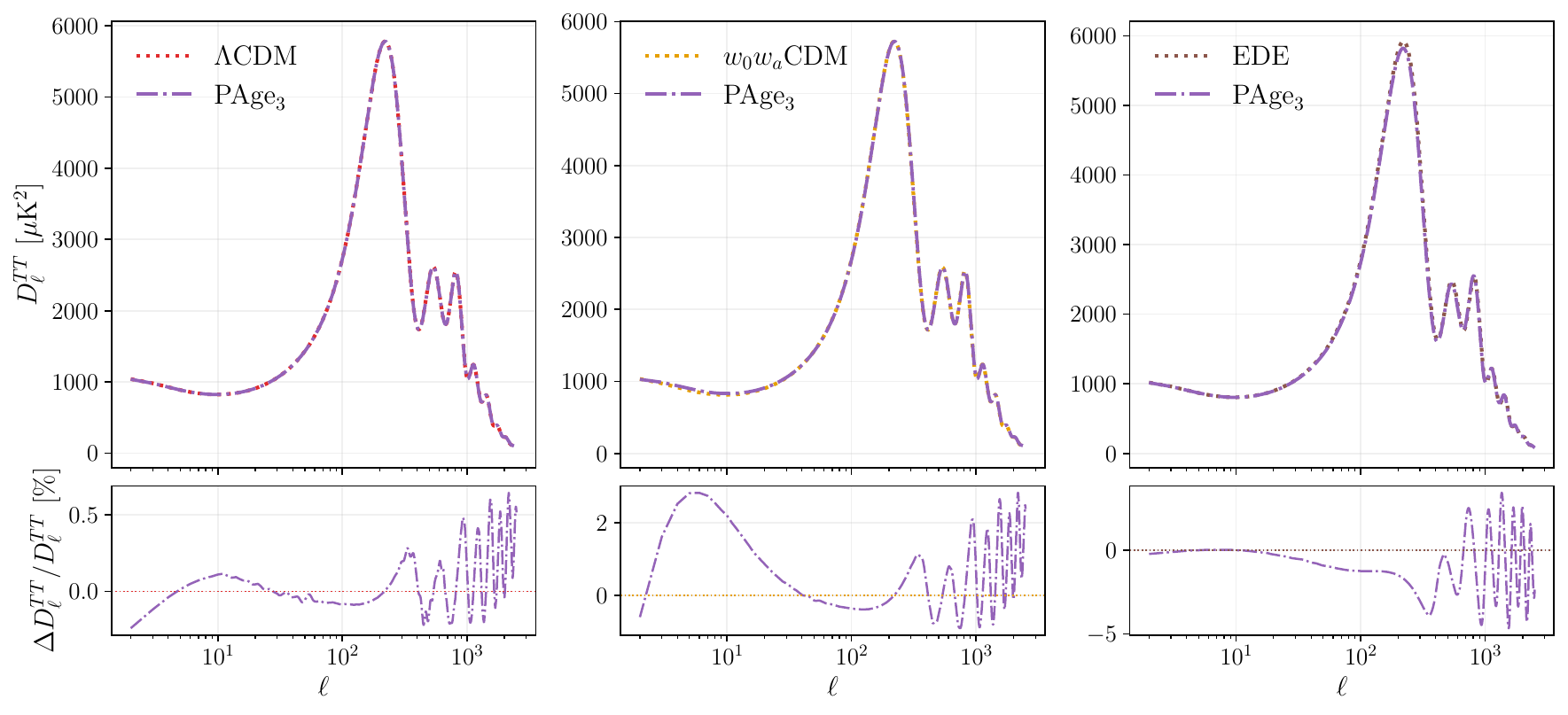}
\caption{CMB TT-spectrum mappings of representative $\Lambda \rm CDM$, $w_0w_a \rm CDM$, and EDE using $\rm PAge_3$. The upper panels show $D_\ell^{ TT}\equiv\ell(\ell+1)C_\ell^{ TT}/(2\pi)$ in $\mu\mathrm{K}^2$, while the lower panels show the corresponding relative residuals.}
\label{fig:cmb-mapping}
\end{figure}

Second, the present-day conformal time inferred from the GPAge trajectory obtained by integrating the background evolution must agree with the input parameter,
\begin{equation}
  y_0=H_0\tau_0=H_0\int_0^1\frac{{\rm d}a}{a^2 H_{\rm PAge}(a)} .
  \label{eq:conformal-time-consistency}
\end{equation}
These two conditions determine $y_*$ and $y_0$ numerically through a two-dimensional shooting procedure. Consequently, the independent GPAge parameters are
\begin{equation}
\left\{p_{\rm age},\eta_1,\eta_2,\eta_3\right\}.
\end{equation}
Fig.~\ref{fig:cmb-mapping} compares the CMB temperature spectra of $\Lambda \rm CDM $, $w_0w_a \rm CDM$ and EDE with their $\rm PAge_3$ mappings. The relative residuals for the $\Lambda \rm CDM$ mapping remain below $\sim1\%$.

\section{Data sets and methodology}\label{sec:data-methodology}

We confront the PAge model with current cosmological observations. The following data sets are employed in our analysis:

\begin{enumerate}

\item \textbf{CMB.}
We combine the temperature, polarization, and CMB-lensing
measurements from Planck and ACT. Specifically, we include
(i) the Planck 2018 low-multipole $(2\le\ell \le30)$ temperature anisotropy power spectrum, $(C^{TT}_{\ell})$, using \textsc{Commander} likelihood~\cite{Planck:2019nip,Planck:2018vyg}. 
(ii) the low-multipole $(2\le\ell \le30)$ E-mode polarization power spectrum $C_{\ell }^{EE}$ using \textsc{SimAll} likelihood~\cite{Planck:2019nip,Planck:2018vyg}.
(iii) the Planck PR4/NPIPE \textsc{CamSpec} high-multipole $(30<\ell)$ power spectra of temperature and polarization (TTTEEE) anisotropies~\cite{Efstathiou:2019mdh,Rosenberg:2022sdy}. 
(iv) the joint ACT DR6 and Planck PR4/NPIPE CMB-lensing likelihood, constructed from the reconstructed lensing potential power spectra~\cite{ACT:2023dou,ACT:2023kun,Carron:2022eyg}.

\item \textbf{DESI.}
We use the DESI Data Release 2 (DR2) BAO measurements from bright galaxies, luminous red galaxies, emission-line galaxies, quasars, and the Ly$\alpha$ forest. The data include the isotropic BAO measurements $D_V(z)/r_{\mathrm{d}}$ and the anisotropic BAO measurements $D_M(z)/r_{\mathrm{d}}$ and $D_H(z)/r_{\mathrm{d}}$, where $r_{\mathrm{d}}\equiv r_{\mathrm{s}}(z_{\mathrm{d}})$ is the drag-epoch sound horizon, as summarized in Table IV of Ref.~\cite{DESI:2025zgx}.

\item \textbf{DES-Dovekie.}
We use the recalibrated and reanalyzed DES Y5 SNe dataset~\cite{DES:2024jxu,Popovic:2025glk,DES:2025sig}. It contains a total of 1820 SNe, including 197 low-redshift SNe and 1623 likely Type Ia SNe with $P_{\rm SNe}>0.5$.

\item \textbf{SH0ES.}
We adopt the local SH0ES measurement $H_0 = 73.04 \pm 1.04~
{\rm km\,s^{-1}\,Mpc^{-1}}$, which is implemented as a Gaussian prior on $H_0$~\cite{Riess:2021jrx}.
\end{enumerate}
We consider two combinations of these data sets, distinguished by whether the SH0ES prior is included,
\begin{equation}
\begin{split}
\mathcal D_0 &={\rm CMB}+{\rm DESI}+{\rm DES\!-\!Dovekie},\\
\mathcal D_1 &= \mathcal D_0+{\rm SH0ES}.
\end{split}
\end{equation}

We compute the theoretical predictions with a modified version of \textsc{CLASS}~\cite{Lesgourgues:2011re,Blas:2011rf}. Cosmological parameter inference is performed with the public \textsc{Cobaya}~\cite{Torrado:2020dgo,Torrado:2019cobaya}. The chains and posterior distributions are analyzed and visualized using \textsc{GetDist}~\cite{Lewis:2019xzd}. The chains are considered converged when the Gelman--Rubin criterion satisfies $R-1<0.05$~\cite{Gelman:1992zz}. The joint likelihood is given by
\begin{equation}
-2\ln\mathcal L_{\rm tot}=\chi^2_{\rm CMB}+\chi^2_{\rm DESI}+\chi^2_{\rm DES-Dovekie}+\chi^2_{\rm SH0ES},
\end{equation}
where the last term is included only for $\mathcal D_1$. For $\mathrm{PAge}_{3}$, the sampled parameters are
\begin{equation}
\boldsymbol{\theta}_3=
\left\{H_0,\omega_{\rm b},\ln(10^{10}A_s),n_s,\tau_{\rm reio}, p_{\rm age},\eta_1,\eta_2\right\}.
\end{equation}
$\mathrm{PAge}_{2}$ fixes $\eta_2=0,\eta_3=0$, $\mathrm{PAge}_{3}$ fixes $\eta_3=0$, while $\mathrm{PAge}_{4}$ additionally samples $\eta_3$. The prior ranges and fixed or derived quantities are summarized in Table~\ref{tab:prior}, where $\mathcal U[a,b]$ denotes a uniform prior on $[a,b]$ and $H_0$ is given in $\mathrm{km\,s^{-1}\,Mpc^{-1}}$.

\begin{table}[!htbp]
    \centering
    \begin{tabular}{cc}
        \hline
        \multicolumn{2}{c}{Cosmological parameters} \\
        \hline
        $\omega_{\rm b}$
        & $\mathcal{U}[0.01,1]$ \\
        $\omega_{\rm cdm}$
        & $\mathcal{U}[0,1]$ \\
        $H_0$
        & $\mathcal{U}[40,110]$ \\
        $\ln(10^{10}A_s)$
        & $\mathcal{U}[2.94,3.24]$ \\
        $n_s$
        & $\mathcal{U}[0.94,1.02]$ \\
        $\tau_{\rm reio}$
        & $\mathcal{U}[0.03,0.12]$ \\
        \hline
    \end{tabular}
    \qquad\qquad
    \begin{tabular}{cc}
        \hline
        \multicolumn{2}{c}{PAge parameters} \\
        \hline
        $\omega_{\rm cdm}$
        & $0$ (fixed) \\
        $p_{\rm age}$
        & $\mathcal{U}[0.4,1.5]$ \\
        $\eta_1$
        & $\mathcal{U}[0,1]$ \\
        $\eta_2$
        & $\mathcal{U}[-0.5,0.5]$ \\
        $\eta_3$
        & $\mathcal{U}[-2,2.5]$ \\
        $H_0\tau_0$
        & Derived \\
        $H_0\tau_*$
        & Derived \\
        \hline
    \end{tabular}
    \caption{Flat priors used in this paper. In $\rm PAge$ models, $\omega_{\rm cdm}=0$, and $H_0\tau_0$ and $H_0\tau_*$ are derived quantities. The $H_0$ values are quoted in $\mathrm{km\,s^{-1}\,Mpc^{-1}}$.}
    \label{tab:prior}
\end{table}

For model comparison, we compare the minimum $\chi^2$ and the Bayesian evidence. We use \textsc{mcevidence}~\cite{Heavens:2017afc} to estimate the logarithmic Bayes factor from the chains,
\begin{equation}
\ln \mathcal{B}_{ij}=\ln Z_i-\ln Z_j ,
\end{equation}
where $Z_i$ and $Z_j$ are the marginal likelihoods of models $M_i$ and $M_j$, respectively. For a model $\mathcal M$, the evidence is defined as
\begin{equation}
Z\equiv P(\mathcal D|\mathcal M)=\int \mathrm{d}\boldsymbol{\theta}\,\mathcal L(\boldsymbol{\theta})\pi(\boldsymbol{\theta}).
\end{equation}
Here, $\boldsymbol{\theta}$ denotes the parameter vector of the model, $\mathcal L(\boldsymbol{\theta})$ is the likelihood, and $\pi(\boldsymbol{\theta})$ is the parameter prior. For descriptive purposes, we use the Jeffreys scale~\cite{Trotta:2008qt} as a guide for model comparison: $|\ln \mathcal{B}_{ij}|<1$ is inconclusive, $1\leq|\ln \mathcal{B}_{ij}|<2.5$ indicates weak evidence, $2.5\leq|\ln \mathcal{B}_{ij}|<5$ moderate evidence, $5\leq|\ln \mathcal{B}_{ij}|<10$ strong evidence, and $|\ln \mathcal{B}_{ij}|\geq10$ decisive evidence. A positive $\ln \mathcal{B}_{ij}$ favors model $i$ over model $j$. In our analysis, $i$ denotes the PAge model and $j$ denotes $\Lambda$CDM. We further use $Q_{\rm DMAP}$ to quantify the tension with SH0ES~\cite{Raveri:2018wln,Schoneberg:2021qvd},
\begin{equation}
  Q_{\rm DMAP}=\sqrt{\chi^2_{\rm min}(\mathcal{D}_1)-\chi^2_{\rm min}(\mathcal{D}_0)}.
  \label{eq:3.6}
\end{equation}

\FloatBarrier
\section{Results and discussion}
\label{sec:results-discussion}

\begin{figure}[!htbp]
  \centering
  \includegraphics[width=0.95\textwidth]{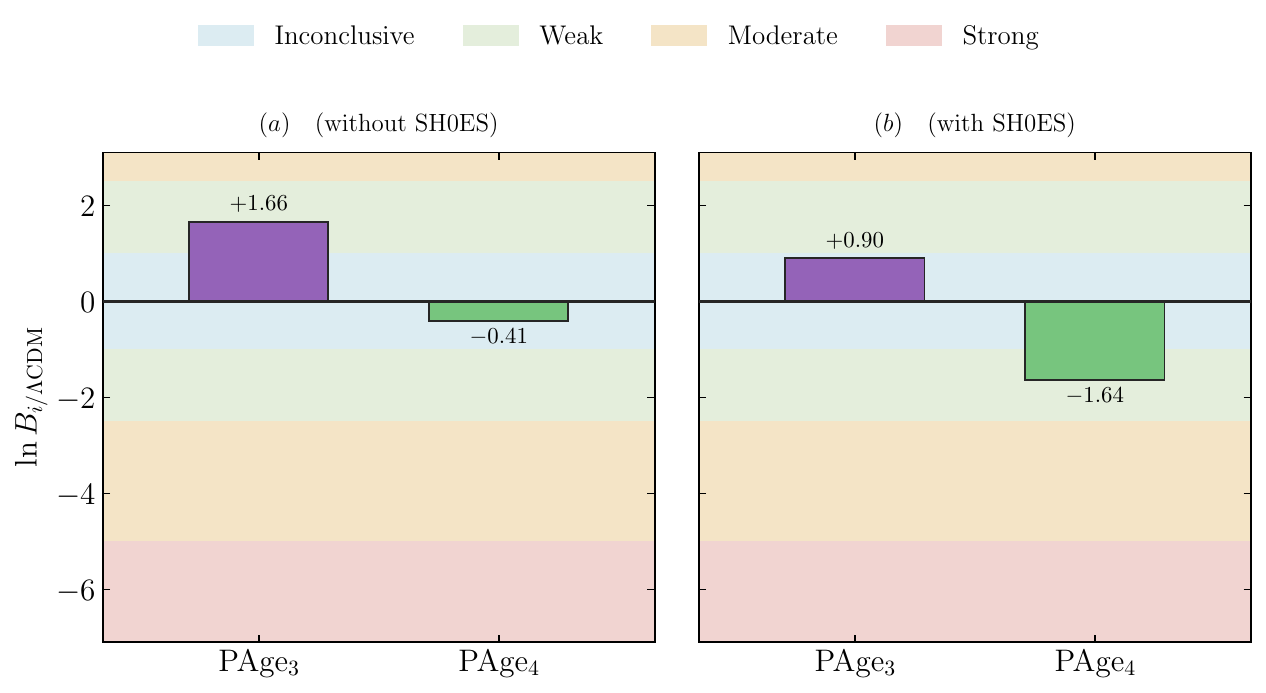}
  \caption{Bayesian model comparison of $\rm PAge_3$ and $\rm PAge_4$
relative to $\Lambda$CDM, without SH0ES (left) and with SH0ES (right).
Positive values indicate support for PAge over $\Lambda$CDM.}
  \label{fig:bayes-comparison}
\end{figure}

\begin{figure}[!htbp]
\centering
\includegraphics[width=1\textwidth]{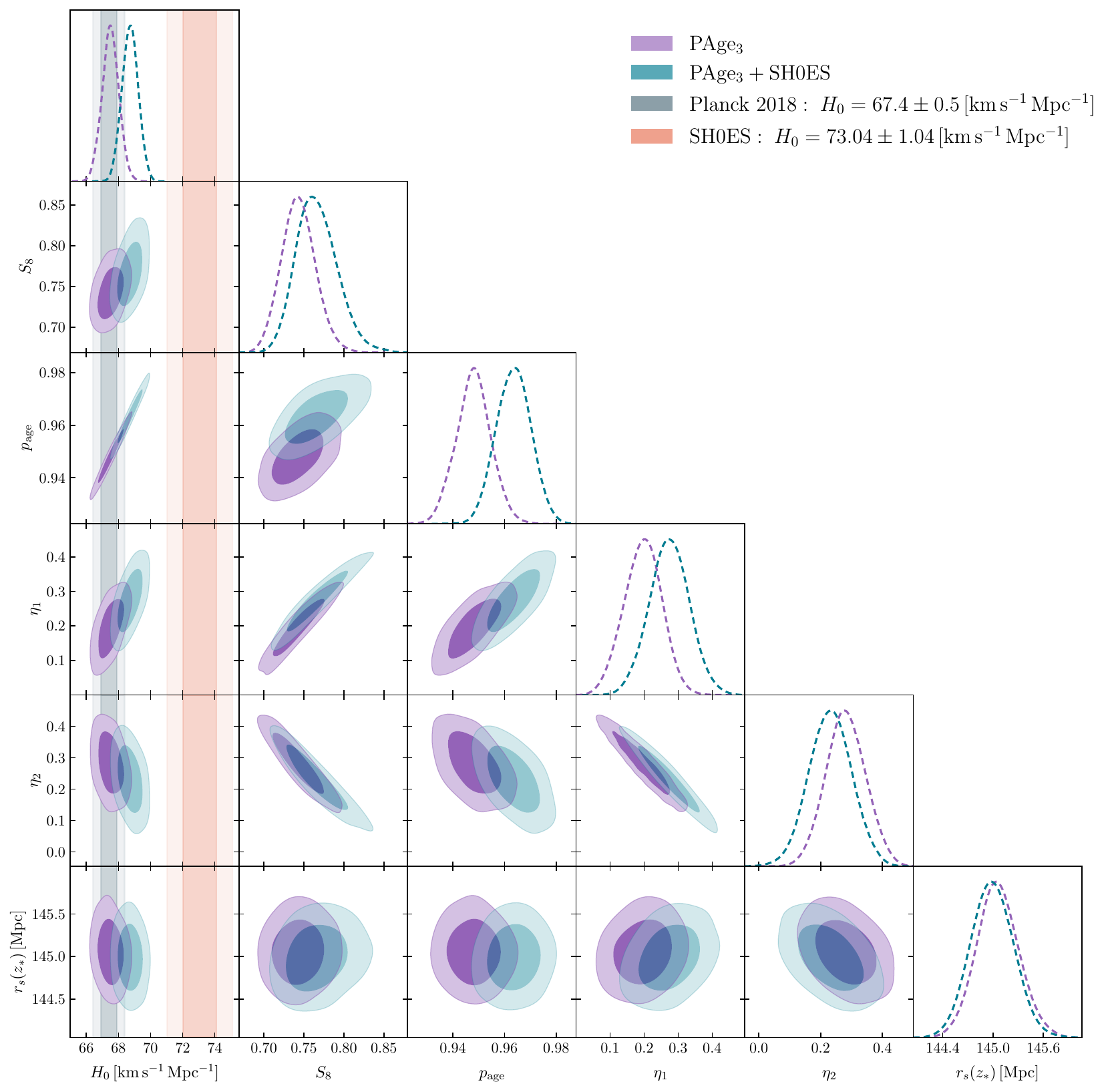}
\caption{Marginalized posterior distributions for $\mathrm{PAge}_3$ with and without the SH0ES prior. Dark and light contours denote the $68\%$ and $95\%$ credible regions, respectively. The grey and salmon bands indicate the Planck 2018 and SH0ES constraints on $H_0$.}
\label{fig:page-posterior}
\end{figure}

\begin{figure}[!htbp]
\centering
\includegraphics[width=0.8\textwidth]{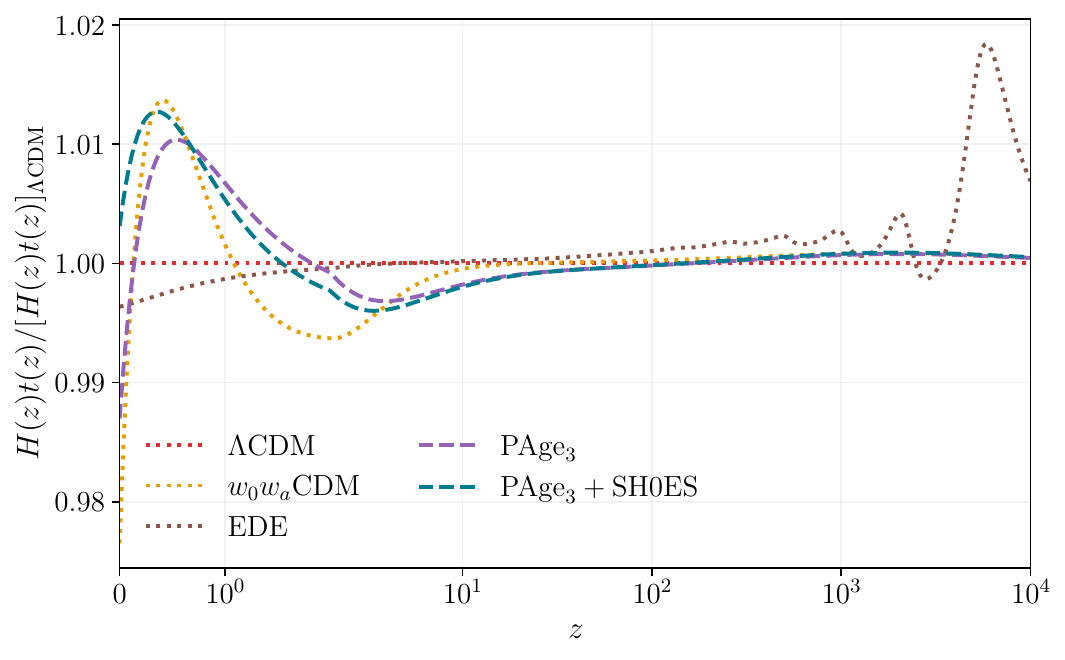}
\caption{$H(z)t(z)$ for the posterior-mean PAge$_3$ model without SH0ES (purple) and with SH0ES (blue), normalized to the \(\mathcal D_0\) posterior-mean $\Lambda \rm CDM$. EDE and $w_0w_a \rm CDM$ are shown for comparison.}
\label{fig:clock-trajectories}
\end{figure}

\begin{figure}[!htbp]
\centering
\includegraphics[width=1.0\textwidth]{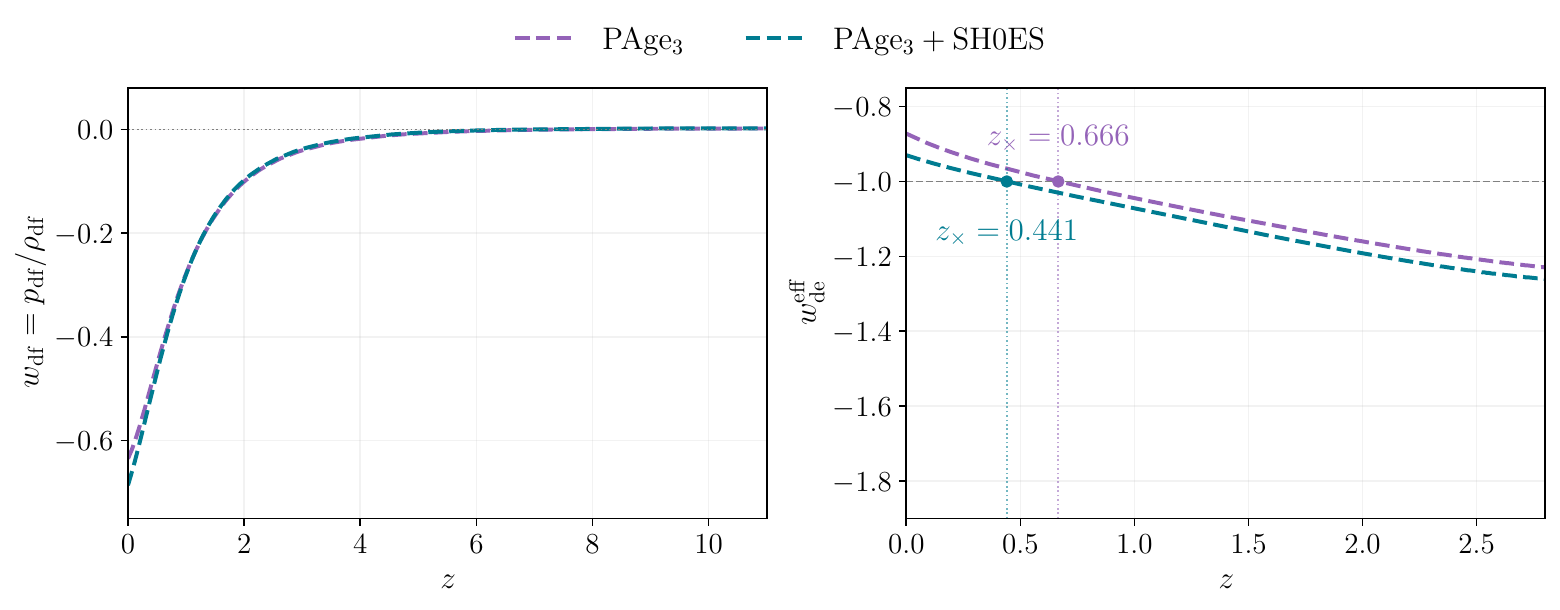}
\caption{EOS of dark fluid and effective DE defined in Eq.~\eqref{eq:udf-background-quantities} and Eq.~\eqref{eq:effective-de-eos} for $\rm PAge_3$ with and without SH0ES. The effective phantom crossings are marked explicitly.}
\label{fig:udf-eos}
\end{figure}

In this section, we first compare the best-fit $\chi^2$ and the Bayes factor for $\Lambda \rm CDM$, $\rm PAge_3$, and $\rm PAge_4$ in Table~\ref{table:bayes_chi2} and Fig.~\ref{fig:bayes-comparison}. We then present the marginalized cosmological constraints for $\rm PAge_3$ in Table~\ref{table:param_values} and Fig.~\ref{fig:page-posterior}. The strengthened no-go result is illustrated in Fig.~\ref{fig:clock-trajectories}, followed by a discussion of the inferred $S_8$ in $\rm PAge_3$. Finally, the effective late-time dynamics and the phantom crossing are discussed in Fig.~\ref{fig:udf-eos}.

\begin{table}[!htbp]

  \centering

  \small

  \setlength{\tabcolsep}{3pt}

  \resizebox{\textwidth}{!}{

  \begin{tabular}{l|c|c|c||c|c|c}

    \hline\hline

Data
    & $\Lambda \rm CDM$
    & $\mathrm{PAge}_{3}$
    & $\mathrm{PAge}_{4}$
    & \shortstack{$\Lambda \rm CDM$\\$+$ SH0ES}
    & \shortstack{$\mathrm{PAge}_{3}$\\$+$ SH0ES}
    & \shortstack{$\mathrm{PAge}_{4}$\\$+$ SH0ES}
    \\

    \hline

    \textit{Planck} high-$\ell$
    & 10547.145
    & 10542.789
    & 10542.183
    & 10547.383
    & 10542.186
    & 10542.361
    \\

    \textit{Planck} low-$\ell$
    & 418.355
    & 419.524
    & 420.026
    & 418.347
    & 419.733
    & 419.485
    \\

    Lensing
    & 20.026
    & 19.953
    & 19.891
    & 20.028
    & 19.964
    & 19.864
    \\

    DESI DR2 BAO
    & 11.618
    & 8.611
    & 8.534
    & 11.529
    & 10.066
    & 10.333
    \\

    DES--Dovekie SNe
    & 1635.289
    & 1630.053
    & 1630.096
    & 1635.342
    & 1631.620
    & 1631.365
    \\

    SH0ES
    & $-$
    & $-$
    & $-$
    & 20.164
    & 18.807
    & 18.751
    \\

    \hline

     $\chi^2_{\min}$
    & 12632.433
    & 12620.930
    & 12620.730
    & 12652.793
    & 12642.377
    & 12642.159
    \\

    $\Delta\chi^2_{\min}$
    & 0.000
    & $-11.503$
    & $-11.703$
    & 0.000
    & $-10.416$
    & $-10.634$
    \\

    $Q_{\rm DMAP}$
    & $-$
    & $-$
    & $-$
    & 4.51
    & 4.63
    & 4.63
    \\

     $\ln \mathcal{B}_{ij}$
    & 0.000
    & $+1.662$
    & $-0.408$
    & 0.000
    & $+0.899$
    & $-1.643$
    \\

    \hline

  \end{tabular}

  }

  \caption{Best-fit $\chi^2$ contributions for $\Lambda \rm CDM$, $\mathrm{PAge}_3$, and $\mathrm{PAge}_4$, with and without the SH0ES constraint. $\Delta\chi^2_{\min}$ and $\ln \mathcal{B}_{ij}$ are defined relative to the corresponding $\Lambda \rm CDM$ model. The Hubble tension is quantified by $Q_{\rm DMAP}$, as defined in Eq.~\eqref{eq:3.6}. PAge$_2$ is not included because it provides a substantially poorer fit to the data, with $\chi^2_{\rm min}=12636.478$.}

  \label{table:bayes_chi2}

\end{table}

Table~\ref{table:bayes_chi2} shows that $\mathrm{PAge}_3$ provides a substantially better fit than $\Lambda \rm CDM$, with $\Delta\chi^2_{\min}=-11.503$ for $\mathcal D_0$ and $\Delta\chi^2_{\min}=-10.416$ for $\mathcal D_1$. For $\mathcal D_0$, the improvement is mainly driven by DESI DR2 BAO, DES--Dovekie SNe, and the Planck high-$\ell$ likelihood. After including SH0ES, PAge retains a substantial improvement in the global fit, although part of the gain in BAO and SN is reduced. The improved fits to BAO and SN are qualitatively consistent with the preference for departures from the $\Lambda \rm CDM$ late-time expansion history found in DESI DR2 analyses~\cite{DESI:2025zgx}.

By contrast, $\mathrm{PAge}_2$ does not provide sufficient freedom to fit the data, yielding $\chi^2_{\min}=12636.478$ for $\mathcal D_0$, and is therefore not considered further. Adding the extra parameter $\eta_3$ in $\mathrm{PAge}_4$ brings only a marginal improvement, lowering $\chi^2_{\min}$ by $0.200$ for $\mathcal D_0$ and $0.218$ for $\mathcal D_1$ relative to $\mathrm{PAge}_3$. The Bayes-factor estimates lead to the same qualitative conclusion. For $\mathcal D_0$, $\mathrm{PAge}_3$ gives $\ln \mathcal{B}_{ij}=1.662$, which is a weak preference relative to $\Lambda \rm CDM$ under the Jeffreys interpretation, whereas the evidence for $\mathrm{PAge}_4$ is inconclusive. With SH0ES included, the evidence for $\mathrm{PAge}_3$ becomes inconclusive, while $\mathrm{PAge}_4$ is weakly disfavored relative to $\Lambda \rm CDM$. These results indicate that the additional freedom introduced in $\mathrm{PAge}_4$ is not supported by the current data, and we therefore focus on $\mathrm{PAge}_3$ in the following analysis.

Despite the improved global fit, the Hubble tension is not alleviated. The $Q_{\rm DMAP}$ statistic increases slightly from $4.51\sigma$ in $\Lambda \rm CDM$ to $4.63\sigma$ in both $\mathrm{PAge}_3$ and $\mathrm{PAge}_4$. The reason why the PAge model cannot reduce the Hubble tension is investigated below. The marginalized constraints for the $\rm PAge_3$ model are summarized in Table~\ref{table:param_values}. For the baseline data sets $\mathcal{D}_0$, we obtain
\begin{equation}
H_0=67.50^{+0.51}_{-0.51}\,\mathrm{km\,s^{-1}\,Mpc^{-1}},
\end{equation}
and
\begin{equation}
r_{\mathrm{s}}(z_*)=145.06^{+0.26}_{-0.25}\,\mathrm{Mpc}.
\end{equation}
Including the SH0ES prior shifts the posterior to
\begin{equation}
H_0=68.73^{+0.49}_{-0.50}\,
\mathrm{km\,s^{-1}\,Mpc^{-1}},
\end{equation}
while the sound horizon at last scattering remains nearly unchanged,
\begin{equation}
r_{\mathrm{s}}(z_*)=144.99^{+0.26}_{-0.26}\,\mathrm{Mpc}.
\end{equation}
Thus, the SH0ES prior shifts $H_0$ upward by approximately $1.2\,\mathrm{km\,s^{-1}\,Mpc^{-1}}$, while producing essentially no shift in $r_{\mathrm{s}}(z_*)$.

\begin{table}[!htbp]
  \centering
  \small
  \setlength{\tabcolsep}{3pt}
  \renewcommand{\arraystretch}{0.9}

  \resizebox{0.82\textwidth}{!}{%
  \begin{tabular}{l|c|c||c|c}
    \hline\hline
    Parameter
    & $\Lambda\mathrm{CDM}$
    & $\mathrm{PAge}_3$
    & \shortstack{$\Lambda\mathrm{CDM}$\\$+$ SH0ES}
    & \shortstack{$\mathrm{PAge}_3$\\$+$ SH0ES}
    \\
    \hline

    $H_0$
    & $68.11_{-0.26}^{+0.29}$
    & $67.50_{-0.51}^{+0.51}$
    & $68.40_{-0.26}^{+0.27}$
    & $68.73_{-0.50}^{+0.49}$
    \\

    $r_{\mathrm{s}}(z_*)$
    & $145.49_{-0.19}^{+0.19}$
    & $145.06_{-0.25}^{+0.26}$
    & $145.57_{-0.15}^{+0.16}$
    & $144.99_{-0.26}^{+0.26}$
    \\

    $\omega_{\rm b}$
    & $0.02232_{-0.00011}^{+0.00012}$
    & $0.02219_{-0.00013}^{+0.00013}$
    & $0.02241_{-0.00012}^{+0.00012}$
    & $0.02219_{-0.00013}^{+0.00013}$
    \\

    $\omega_{\rm cdm}$
    & $0.1179_{-0.0006}^{+0.0006}$
    & $-$
    & $0.1173_{-0.0005}^{+0.0005}$
    & $-$
    \\

    $\ln(10^{10}A_s)$
    & $3.055_{-0.013}^{+0.013}$
    & $3.045_{-0.013}^{+0.013}$
    & $3.057_{-0.013}^{+0.013}$
    & $3.045_{-0.013}^{+0.013}$
    \\

    $n_s$
    & $0.9686_{-0.0036}^{+0.0035}$
    & $0.9640_{-0.0038}^{+0.0040}$
    & $0.9703_{-0.0032}^{+0.0032}$
    & $0.9633_{-0.0034}^{+0.0036}$
    \\

    $\tau_{\rm reio}$
    & $0.0608_{-0.0074}^{+0.0073}$
    & $0.0546_{-0.0071}^{+0.0070}$
    & $0.0628_{-0.0077}^{+0.0068}$
    & $0.0544_{-0.0072}^{+0.0067}$
    \\

    $p_{\rm age}$
    & $-$
    & $0.9480_{-0.0068}^{+0.0065}$
    & $-$
    & $0.9635_{-0.0068}^{+0.0066}$
    \\

    $\eta_1$
    & $-$
    & $0.196_{-0.053}^{+0.057}$
    & $-$
    & $0.274_{-0.058}^{+0.057}$
    \\

    $\eta_2$
    & $-$
    & $0.282_{-0.063}^{+0.063}$
    & $-$
    & $0.232_{-0.068}^{+0.069}$
    \\

    \hline

    $S_8$
    & $0.8139_{-0.0068}^{+0.0071}$
    & $0.743_{-0.021}^{+0.020}$
    & $0.8088_{-0.0072}^{+0.0068}$
    & $0.766_{-0.028}^{+0.022}$
    \\

    \hline
  \end{tabular}%
  }

  \caption{Marginalized posterior means and $68\%$ credible intervals for $\Lambda\mathrm{CDM}$ and $\mathrm{PAge}_3$, with and without the SH0ES $H_0$ prior. $H_0$ and $r_{\mathrm{s}}(z_*)$ are in units of $\mathrm{km\,s^{-1}\,Mpc^{-1}}$ and $\mathrm{Mpc}$, respectively. Here $r_{\mathrm{s}}(z_*)$ denotes the sound horizon at last scattering.}
  \label{table:param_values}
\end{table}

The correlated parameter shifts are shown in Fig.~\ref{fig:page-posterior}. Adding SH0ES shifts $H_0$ together with $p_{\rm age}$, $\eta_1$, and $\eta_2$ along the allowed PAge parameter directions, but does not resolve the Hubble tension. The small change in $r_{\rm s}(z_*)$ reveals the limited early-time freedom of the PAge parametrization. Relative to $\Lambda$CDM, PAge$_3$ lowers $r_{\rm s}(z_*)$ by only about $0.3$--$0.4\%$ for both $\mathcal D_0$ and $\mathcal D_1$. As shown in Eq.~\eqref{eq:2.28}, once the radiation abundance is fixed, $y_*$ mainly controls the early-time matter density and hence shifts matter--radiation equality, which is tightly constrained by the CMB. The remaining PAge freedom mainly changes the late-time expansion and therefore has little leverage on $r_{\rm s}(z_*)$. This limitation is further illustrated in Fig.~\ref{fig:clock-trajectories}. The preferred PAge$_3$ trajectories remain smooth, whereas EDE introduces a localized pre-recombination feature around matter--radiation equality.

An additional consequence of the UDF description is a lower inferred clustering amplitude. We obtain
\begin{equation}
S_8=0.743^{+0.020}_{-0.021}\quad\text{for }\mathcal D_0,
\end{equation}
and
\begin{equation}
S_8=0.766^{+0.022}_{-0.028}\quad\text{for }\mathcal D_1,
\end{equation}
both below the corresponding $\Lambda \rm CDM$ values. Here, $S_8$ is defined using the effective clustering matter density in Eq.~\eqref{eq:2.25}, rather than a separately conserved CDM component. Its relatively low value therefore reflects, in part, the lower effective clustering matter density entering the definition of $S_8$. This feature is closely related to interacting dark-sector models, where energy exchange can change the clustering matter abundance and suppress late-time structure growth, thereby reducing $S_8$~\cite{Khoury:2025txd}.

We finally examine the effective late-time dark-energy dynamics using Eqs.~\eqref{eq:udf-background-quantities} and \eqref{eq:effective-de-eos}.
As shown in Fig.~\ref{fig:udf-eos}, the UDF EOS $w_{\rm df}$ remains above $-1$, whereas the effective dark-energy equation of state exhibits a phantom crossing. The crossing occurs at $z_\times=0.666$ for $\mathcal D_0$ and shifts to $z_\times=0.441$ after including SH0ES. This shift reflects the response of the effective late-time dynamics to the SH0ES prior.

\FloatBarrier

\section{Conclusions}\label{sec4}

The over $5\sigma$ Hubble tension remains one of the major problems in modern cosmology. Proposed resolutions can be broadly classified into early- and late-time solutions, but no compelling consensus has yet emerged. This motivates phenomenological parametrizations that can explore a broad class of possibilities with minimal assumptions. While late-time parametrizations have been extensively studied, extending such approaches to the early Universe is more challenging because the CMB constrains both the background evolution and the perturbation sector. A framework capable of parametrizing both early- and late-time modifications is therefore valuable for investigating the Hubble tension.

In this work, we extend the original PAge parametrization from the matter-dominated regime to the radiation-dominated epoch. The original PAge construction starts from the matter-era relation $Ht=2/3$. We replace this baseline with the exact radiation--matter solution
\begin{equation}
  Ht=\frac{2}{3}\frac{(\tau/\tau_*+1)(\tau/\tau_*+3)}{(\tau/\tau_*+2)^2},
  \label{eq:5.1}
\end{equation}
which smoothly connects the radiation-dominated limit $Ht=1/2$ to the matter-dominated limit $Ht=2/3$. To confront the model with the CMB, we complete the perturbation sector within a unified-dark-fluid framework with $c_{\rm s,eff}^2=0$ and vanishing anisotropic stress.

Confronting this parametrization with CMB+DESI DR2+DES-Dovekie, with and without SH0ES, we find that PAge$_3$ improves the best fit relative to $\Lambda$CDM. However, the Hubble tension is not alleviated, while $r_{\rm s}$ remains essentially unchanged. This reflects the limited early-time freedom of the parametrization. Our results show that retuning the early-time matter density, even when combined with a flexible late-time expansion history, remains insufficient to resolve the Hubble tension. Additional pre-recombination physics is therefore required.

The UDF interpretation also leads to a substantially lower effective clustering amplitude. Since the effective clustering matter component is not separately conserved, the inferred $S_8$ should be interpreted as an effective clustering measure rather than directly identified with the standard CDM-based quantity. This behavior is qualitatively similar to that found in interacting dark-sector models, where the evolution of the clustering matter component can suppress late-time structure growth. The UDF further produces a phantom crossing in the effective dark-energy equation of state, while the physical UDF itself remains nonphantom.

Future parametrizations of both early- and late-time solutions should account for both the background and perturbation sectors. At the background level, additional freedom should be introduced into Eq.~\eqref{eq:5.1} to capture the localized pre-recombination feature characteristic of early-time solutions. At the perturbation level, additional freedom is also required so that sizable early-time background modifications can remain compatible with the CMB anisotropy spectra. Constructing such a joint background--perturbation parametrization provides a natural next step for extending the PAge framework.
 
\begin{acknowledgments}
This work is supported by the National Key Research and Development Program of China Grant No. 2021YFC2203004, No. 2021YFA0718304, and No. 2020YFC2201501, the National Natural Science Foundation of China Grants No. 12475067, No. 12422502, No. 12105344, No. 12235019, No. 12447101, No. 12073088, No. 11821505, No. 11991052, and No. 11947302, and the China Manned Space Program Grant No. CMS-CSST-2025-A01. We also acknowledge the use of the HPC Cluster of ITP-CAS.
\end{acknowledgments}

\bibliography{main}
\bibliographystyle{JHEP}
\end{document}